# RADAR Imaging in the Open field
# At 300 MHz-3000 MHz Radio Band

Rabindranath Bera [1], Jitendranath Bera[2], Sanjib Sil [3],
Sourav Dhar[1], Debdatta Kandar [4], Dipak Mondal[1]

[1]Sikkim Manipal Institute of Technology, Sikkim Manipal University,
Majitar, Rangpo, East Sikkim 737132, India
Phone: 03592-246220 ext 280; Fax: 03592-246112
E-mail : r.bera @rediffmail.com

[2]Department of Applied Physics, University of Calcutta
92 Acharya Prafulla Chandra Road, Kolkata 700 009, India

[3]Institute of Radio physics & Electronics, University of Calcutta
92 Acharya Prafulla Chandra Road, Kolkata 700 009, India.

[4]Department of Electronics & Telecommunication Engineering
Jadavpur University, Kolkata 700 032, India

*Abstract*

*With the technological growth of broadband wireless technology like CDMA and UWB , a lots of development efforts towards wireless communication system and Imaging radar system are well justified. Efforts are also being imparted towards a Convergence Technology.. the convergence between a communication and radar technology which will result in ITS ( Intelligent Transport System ) and other applications. This encourages present authors for this development. They are trying to utilize or converge the communication technologies towards radar and to achieve the Interference free and clutter free quality remote images of targets using DS-UWB wireless technology.*

*Key Words: CDMA, UWB, DS-UWB, ITS, Stealth ness*

.
## Introduction
**Past Experience and problem Definition**

The authors' past experience of development of an Imaging radar Instrumentation system at W-band over last 10 years ia a closed chamber reminds them the following points:-
1. Narrow band operation of network Analyzer (bandwidth 10 KHz) used for radar Instrumentation restricts the image resolution.
2. Step frequency waveform over a bandwidth of 2 GHz center around 94 GHz and using 201 acquisition points improves the range resolution considerably. But the system becomes sluggish ( typically a time of 20 minutes is required to complete a single measurement).
3. Clutter effects outside the Quiet zone of the closed chamber is severe which results in some clutter energy distribution over the quite zone of the chamber.
4. This ultimately introduces extra Phase Noise in the system resulting Weak target Spot measurement sometimes unstable, particularly for small targets. For example , the radar return signal of a small sphere (used for the calibration of radar) -30 DBSM RCS value was found to be oscillatory from sweep to sweep and very difficult to predict the actual RCS value.

## Caution

In the Imaging Radar Instrumentation system in the Open field , particularly the same clutter problem may become more serious and should not be allowed to repeat. Additionally, the Active Interference from Operational Radio system

( mostly lies in the 300 – 3000 MHZ Radio band) and man made noise are severe. So a suitable modern radar technology is to be justified which will be able to tackle all above problems.

## Necessity of Broadband Operarion

The vast majority of modern radio systems has a narrow frequency range and carrier waveform uses harmonic (sinusoidal) or quasi harmonic signals to transmit information. The reason is very simple- sinusoidal oscillations are generated by the RLC oscillation contour itself. The narrow frequency range of the signal restricts the informational capacity of the radio system. So it is necessary to expand the frequency range in order to increase the information capacity. Future development of radar lies in employment of signals with frequency range up to 1 GHz (the duration of the radiated pulses around 1 ns). The informational content in the UWB location increases owing to the range reduction of the pulse volume of radar. Thus, when the length of a pulse changes from 1 µs to 1 ns the depth of the pulse volume reduces from 300 m to 30 cm.

It could be said that the instrument, which investigates the space, becomes more fine and sensitive. It allows to obtain the radio image of the targets[1].

## The Target Scattering Cross Section for UWB signals

UWB is a recent radio technique, specification of which is already formulated by IEEE in the form of IEEE802.15.3 standard.

## Stealth ness of an aircraft[2]

The answer to the question 'what's the use of measurement of Radar Cross Section of an aircraft?' may be the determination of Stealth ness of the aircraft so that the particular aircraft may not be detected/recognized by the enemy. The stealth ness varies with RCS values.

Let us consider the difference between reflected signals when a target is irradiated by narrowband and UWB signals. The spatial physical length of a narrowband signal is equal to $c\tau_{NB} \geq L$, where L is the size of a target and $c\tau_{NB} \leq L$ for an UWB signal. The 'long' narrowband signal reflected from all N brilliant points will present a sum of N arbitrary phase harmonic oscillations or their vector sum. In this case, the target scattering cross section is equal to

$$\sigma_{NB} = \sum \sigma_k \cos(2\Pi R/\lambda) \qquad (1)$$

where $\sigma_k$ = the scattering cross section of a brilliant point number k.

R = the distance from the radar to this point.

We have a different picture when the UWB signal, which has $c\tau_{NB} < L$, is reflected from a target. Video pulses making the whole image may have different amplitude. It depends on the scattering cross section of the corresponding brilliant point of the target. The polarity of these pulses may change. As a result, the target RCS becomes time dependent. If the UWB signal processing algorithm allows for adding the reflections from individual brilliant point, then the target RCS is not time dependent.

$$\sigma_{UWB} = \sum \sigma_k \qquad (2)$$

Thus $\sigma_{UWB} > \sigma_{NB}$ i.e. the UWB signal provides a gain in the RCS magnitude.

So the stealth ness of an aircraft varies with the type of radar used for its measurement and a stealth aircraft may become visible to an UWB radar. It is then always better to use the latest radar technology for an Imaging Instrumentation System.

## Work Definition

Our development efforts of radar system can be categorized into two types
1. Spread Spectrum based CDMA Approach using sinusoidal oscillator as carrier
2. Ultra Wide Band (UWB) Approach using Impulse Transmission [3]

## Spread Spectrum Based CDMA Approach [4]

### Salient Features:
* Spread spectrum based digital technology utilized for better radar operation
* Provision for both DSSS & FHSS mode of operation
* QPSK modulation is utilized for better spectral efficiency.
* Spreading codes enhances security as well as

solves lots of radio wave propagation problems and can be changed any time.

* Pulse mode provides range resolution and also

hardware gating of the RF carrier.

* Measurements supports VV, HH, VH, HV Polarization
* Full Automation through PC programming in Instrument settings, Control, Data Acquisition & Display.

A Digital radar, schematic of which is shown in fig.1, is configured utilizing DSSS & FHSS technology in the VHF/UHF band. The Spread spectrum generator is the heart of this Instrumentation radar and is a highly versatile Instrument. Several unique features like support of both DSSS & FHSS mode make it ideal for OPEN RANGE Imaging system.

*QPSK Transmitter*

In the configuration shown in fig. 1., the $I_{DS}$ & $Q_{DS}$ outputs are coming out from the SS generator and are used to modulate two RF carrier signals that are offset in phase by 90 degree. The QPSK modulated signals $I_M$ & $Q_M$ thus formed, are then combined using a combiner to produce a composite spread spectrum signal. ($I_{DS}$ & $Q_{DS}$ are the DSSS modulated analog output and $I_{PN}$ is the digital PN sequence output from the generator). The RF carrier can be synthesized through a PC using its GPIB port control to cover the full transmitter frequency range of 300-3000 MHz. The composite RF signal thus formed is then amplified using solid-state power amplifier to produce the high power RF signal. This is then chopped using a PIN switch controlled by a pulse generator to produce the Pulsed RF Waveform. The range resolution will be governed by the pulse width and unambiguous range extent by the Pulse repetition. For polarization switching, a polarization switch is incorporated which have both vertical & horizontal modes of operation. The antenna with Cassegrain feed is utilized as the transmitting element. All the Major Instruments like SS Generator, Synthesizer and pulse generator have GPIB interface which is being controlled through the PC.

**Receiver**

The LNA output of the received signal from another Cassegrain antenna is fed to the PIN switch, which is operated through the other channel of the pulse generator. The PIN switch at the front end of both the Tx & Rx module is provided to avoid the direct path reception of the transmitted signal. The output of the receiver PIN switch is down converted to IF at 70 MHz with the help of a synthesized PN modulated LO in the range of 200-4000 MHz. The synthesized PN modulated LO is produced with the help of an up converter whose inputs are $I_{PN}$ output of the SS generator and another synthesizer (200-4000 MHz). So, to de-scramble the PN coded IF carrier, the same transmitted PN code coming out from $I_{PN}$ output of the SS generator is used as per the protocol of the SS technology. As a result the composite RF signal collapses back to the original data bandwidth. The collapsed signals $I_C$ & $Q_C$ at 70 MHz is then demodulated with the help of QPSK demodulator which ultimately feeds the signal to the PC for its processing. The PC software processes the data for the object's information in the form of images.

**Ultra Wide Band ( UWB) Approach using Impulse Transmission**

Figure 2 shows the block diagram of the UWB Transmitter with its major components. The pulse generator of the transmitter generates a mono-cycle pulse of 0.33 non-second pulse width. The Code in the form of data input is fed to the programmable Time Delay circuit where it produces Direct Sequence Spread spectrum signal ( DSSS). This DS signal is further used to modulate the Pulse so that the resulting signal will become **DS-UWB signal**. The pulse is amplified by the power amplifier and is radiated by the transmitting antenna.

Figure 3 shows the block diagram of the UWB receiver. The reflected signals from the target will be received by the receiving antenna and is routed through a Correlator front head to produce a low-frequency signal. This low frequency signal, containing information of the targets, is then amplified by the amplifier and after passing through the Sample & Hold circuits it is fed to the Base band Processor.

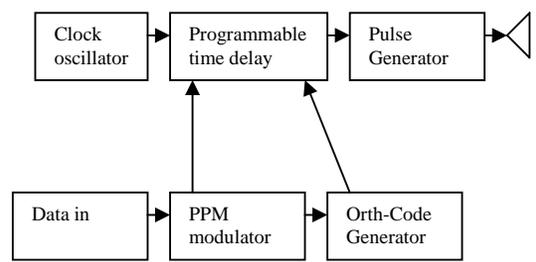

Fig.2: Block diagram of the UWB Transmitter

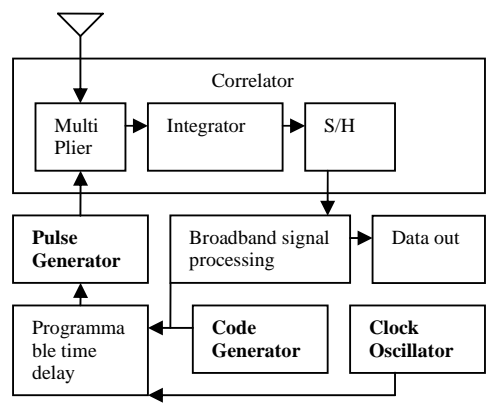

Fig.3: Block Diagram of UWB Receiver

### Authors Information

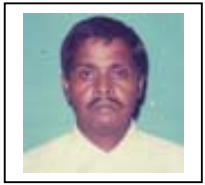
Rabindranath Bera: Born in 1958 at Kolaghat , West Bengal, B. Tech, M. Tech & Ph.D (Tech) from the Institute of Radiophysics & Electronics, The University of Calcutta, in the year 1982,1985 & 1997 respectively. Currently working as Professor and Head of the Deparment, Electronics & Communication Engineering, Sikkim Manipal University, Sikkim, Microwave/ Millimeter wave based Broadband Wireless Mobile Communication and Remote Sensing are the area of specialisatoion.

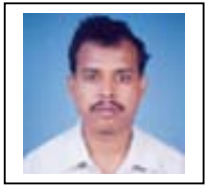
Jtendranath Bera: Born in 1969 at Sajinagachi, West Bengal, B. Tech, M. Tech from the Dept. of Applied Physics, The University of Calcutta, in the year 1993,1995 respectively, and Ph.D.( Engg.) from Jadavpur University, Kolkata. In the year 2005. Currently working as Reader, Dept. of Applied Physics, The University of Calcutta. Microwave/ Millimeter wave based Broadband Wireless Mobile Communication, Remote Sensing and Embedded System are the area of specialisatoion.

Sanjib Sil: Born in 1965 at Kolkata, West Bengal, B. Tech from IETE ( 1989), M. Tech from BIT, Meshra (1991). Ph.D. registration from the Institute of Radiophysics & Electronics, The University of Calcutta ( 2002). Currently working as Asst. Professor, International Institute of Information Technology , Kolkata. Microwave/ Millimeter wave based Broadband Wireless Mobile Communication, Remote Sensing are the area of specialisatoion.

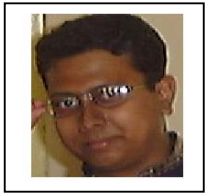
Sourav Dhar: Born in 1980 at Raiganj, West Bengal, B. E from Bangalore Institute of Technology, Visveswaraiah Technological University in the year 2002, M. Tech from Sikkim Manipal Instiutte Of Technology, Sikkim Manipal University in the year 2005. Currently working as Lecturer, Dept. of Electrical & Electronics, SMIT, Broadband Wireless Mobile Communication is the area of specialisatoion.

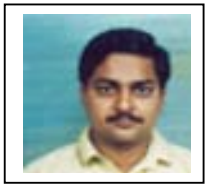
Debdatta Kandar: Born in 1977 at Deulia, West Bengal, B.Sc. ( Honours) from The University of Calcutta in the year 1997, M. Sc from Vidyasagar University in the year 2001. Currently working as Research Fellow in the dept. of Electronics & Telecommunication, Jadavpur University. Mobile Computing and Remote Sensing are the area of specialisatoion.

Dipak Mondal: Born in 1976 at Baruipur, West Bengal, B. Tech, M. Tech from the Institute of Radiophysics & Electronics, the University of Calcutta, in the year 2002, 2004 respectively. Currently working as Lecturer, Dept. of Electronics & Comm. Engg., Sikkim Manipal University, Microwave/ Millimeter wave based Broadband Wireless Mobile Communication, Remote Sensing are the area of specialisatoion.

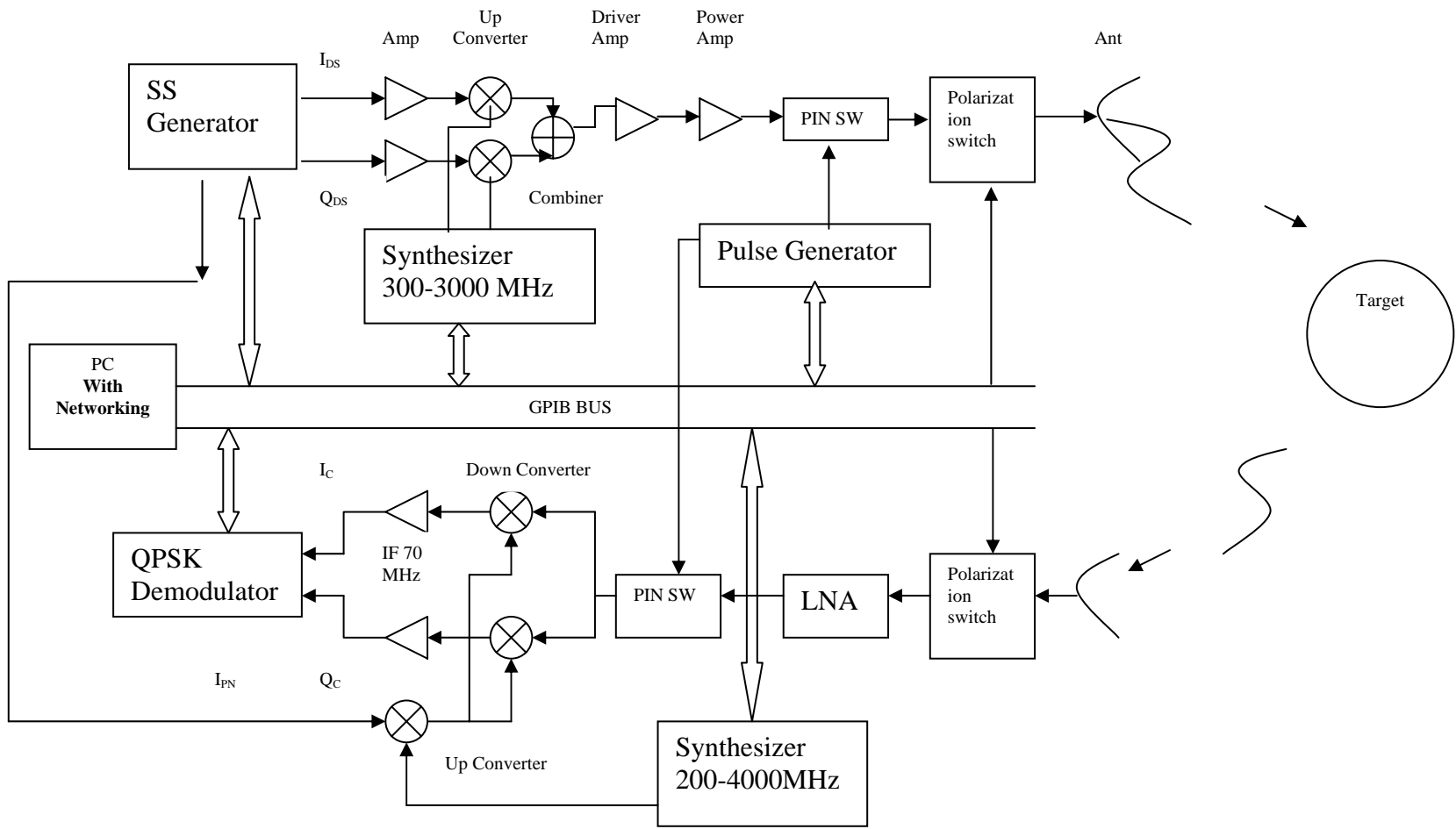

Fig.1. Block diagram of Open Range RCS Measurement RADAR System